\documentclass[12pt]{article}

\usepackage{graphics}
\usepackage[dvips]{graphicx}

\begin{document}

{\footnotesize jcis@epacis.org}

\begin{center}

{\bf OGCOSMO: An auxiliary tool for the study of the Universe within hierarchical scenario of structure formation}
\bigskip


{\small Eduardo S. Pereira\footnote{duducosmo@das.inpe.br}, Oswaldo D. Miranda \footnote{oswaldo@das.inpe.br}
}
\smallskip

{\small
 National Institute for Space Research, S\~ao Jos\'e dos Campos, SP, Brazil
}

{\footnotesize Received on September , 2010 / accepted on *****, 2010}

\end{center}


\begin{abstract}

In this work is presented the software OGCOSMO. This program was written using high level design methodology (HLDM), that is based on the use of very high level (VHL) programing language as  main, and the use of the intermediate level (IL) language only for the critical processing time. The languages used are PYTHON (VHL) and  FORTRAN (IL). The core of OGCOSMO is a package called OGC{\_}lib. This package contains a group of modules for the study of cosmological and astrophysical processes, such as: comoving distance, relation between redshift and time, cosmic star formation rate, number density of dark matter haloes and mass function of supermassive black holes (SMBHs). The software is under development and some new features will be implemented for the research of stochastic  background of gravitational waves (GWs) generated by: stellar collapse to form black holes, binary systems of SMBHs. Even more, we show  that the use of HLDM with PYTHON and FORTRAN is a powerful tool for producing astrophysical softwares.

\bigskip

{\footnotesize
{\bf Keywords}:Computational Physics, Cosmology, Gravitational Waves, Black Hole, Structure Formation.}
\end{abstract}

\textbf{1. INTRODUCTION}
\bigskip
\bigskip

Around  300 thousand years after the Big Bang, matter and radiation decoupled and the photons could freely travel through the space. Henceforth, today we can obtain information about this period from the Cosmic Microwave Background Radiation (CMBR). Moreover, about 840 million years after the Big Bang, the Universe was reionized by the first stars. However, we have no direct information from the period between 300 thousand years and 840 million years. This period is so-called dark age (DA). On the other hand, the theory of general relativity predicts the existence of gravitational waves (GWs) as perturbations of space-time which propagate at the speed of light. The detection of GWs will open a new astronomical window for observing the Universe. In particular, they will allow us a deeper understand about the DA.

In this context, we are interested in to describe, by analytical and semi-analytical models, some astrophysical and cosmological processes as, for example, the cosmic star formation rate, and its connection with the stochastic background of GWs \cite{pereira}, this kind of study will shed light on the knowledge of the process that took place at the end of the DA. 

As such, this modeling framework will provide us with an important auxiliary tool for the reconstruction of the cosmic history. Thus, we decided to organize the computational development as a software, it called OGCOSMO, whose main characteristics are discussed in this work.

In section 2 it will be showed a short description of the astrophysical and cosmological model adopted. In section 3 it will be presented the main characteristics of the software and the type of programing used. In section 4 we present some results. Finally, in section 5 are the final considerations. 

\bigskip
\bigskip

\textbf{2. The Model}

\bigskip
\bigskip

In this work we are considering the general theory of relativity with cold dark matter and cosmological constant ($\Lambda$CDM model) for background cosmology. 

Here is assumed the hierarchical scenario of structure formation, having as base the Press-Schechter (PS) like formalism \cite{PS}. The basic idea behind this scenario is that the formation of objects like galaxies and galaxy clusters occur in the following way: First, when the mean density of dark matter perturbation, within a given volume, is larger than a threshold level, $\delta_{c}$, the matter leaves the linear regime and collapses to form  small-halo objects. These halos become gravitational wells that attract  the baryonic matter, that is the ordinary matter that form stars and planets. Durant all this process, greater halos are formed by fusion of the minors and more baryons fall into structures. The star formation starts and black holes grow up by accretion of matter. The complete details, about the all considerations and the main results obtained with this model, can be seen in \cite{pereira,PereiraB:2010}.

\bigskip
\bigskip

\textbf{3. The Software}

\bigskip
\bigskip

The OGCOSMO software was written using object-oriented programing paradigm, that permits to construct codes that are really reusable and clean. The  main programming language used was PYTHON \cite{pythonof}, that is a very-high level language programing,  and FORTRAN only for the critical time parts of the code. This form of writing codes is called of high level design \cite{hinsenetal,kh}. The external modules used were basically: Tkinter \cite{tkinter} for construction of the Graphical Using Interface (GUI), scipy and  numpy for numerical methods \cite{scipy}.

The core of the OGCOSMO is the OGC{\_}lib, that is a package containing the following principal modules, up to now:

\begin{itemize}
\item OGC{\_}cosmo: This module contains the class ``Cosmo()'' that is composed by the  cosmological background  methods, and callbacks, that are, comoving distance and comoving volume, relation between time and redshift ($z$), matter density evolution (dark and baryonic matters), grow function of matter perturbation,  variance of matter linear density field, linear extrapolation of the critical density for collapse of structures in a given $z$ , comoving volume of dark matter halo.

\item OGC{\_}PS: This module contains the class ``PressSchechter(Cosmo)'', that inherit the methods of Cosmo class. This class contains the base of Press-Schechter like formalism, that is, the functions used for study of structure formation as, mass function and numerical density of dark haloes, fraction and accretion rate of baryonic matter within structures and the cosmic star formation rate - CSFR.

\item OGC{\_}SMBH: In this module is the ``SuperBH(PressSchechter)'' class, that is a class under development and it will be used for the study of the evolution of supermassive black holes  through of its mass function and by the coalescence of these objects \cite{PereiraB:2010}.
\end{itemize}

\bigskip
\bigskip

\textbf{4. Results}



\bigskip
\bigskip
\textbf{4.1 The OGCOSMO}
\bigskip

\begin{figure}[h!]
	\begin{center}
		{\resizebox{0.8\columnwidth}{!}{\includegraphics{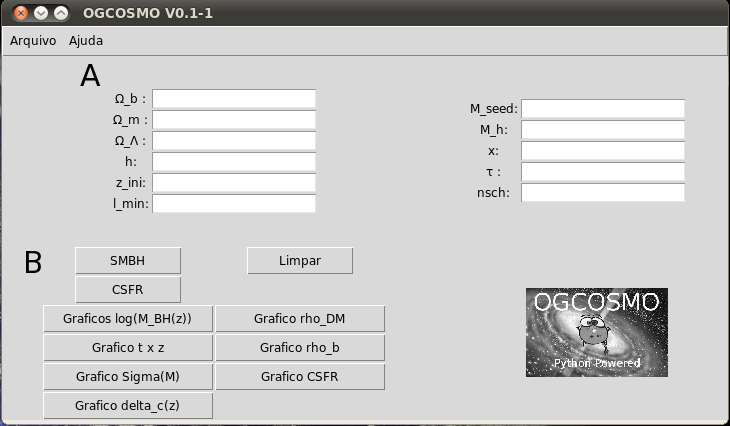}}}
	\end{center}
	\caption{GUI of OGCOSMO. In A are showed the entry parameters. In B are showed the buttons for running the models and generate some graphics.}
	\label{fig:1}
\end{figure}

\begin{figure}[h!]
	\begin{center}
		{\resizebox{0.5\columnwidth}{!}{\includegraphics{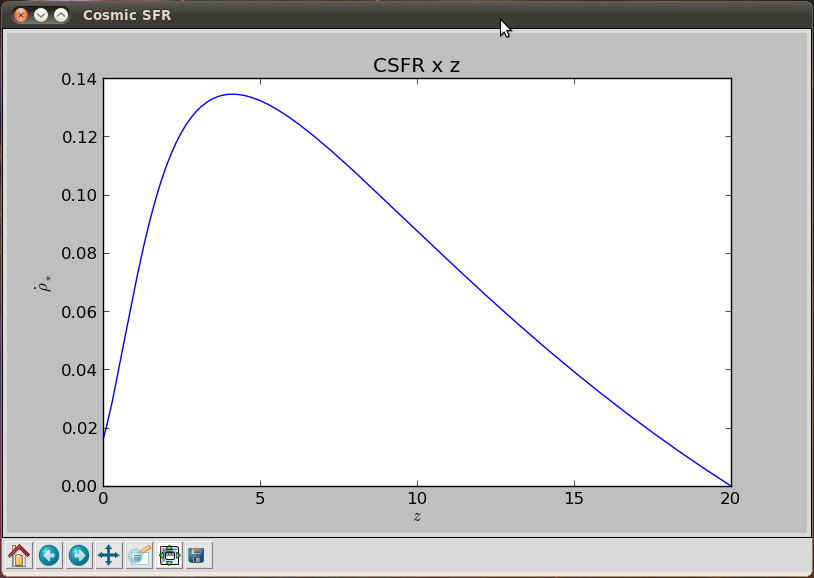}}}
	\end{center}
	\caption{Plot of the cosmic star formation rate behavior. In this case was considered $\Omega_{m} = 0.24$, $\Omega_{b} = 0.04$, $\Omega_{\Lambda} = 0.76$ and $h = 0.73$, for the cosmological parameters, $x=1.35$ and $\tau=2.5\times 10^{9}$ years. }
	\label{fig:2}
\end{figure}

In the figure \ref{fig:1} we show the GUI of OGCOSMO. In A are the spaces for entry parameters such as the cosmological and those associated with the star formation rate. In B are the buttons that start the calculus of the CSFR, supermassive black holes mass function and some buttons for creation of specific plots as the CSFR (see figure \ref{fig:2}).

In the figure \ref{fig:2} is represented the CSFR behavior, whose result can be seen by pressing the button ``Grafico CSFR''. The Salpeter exponent of initial mass function  is  $x=1.35$, the characteristic scale for star formation is $\tau=2.5\times 10^{9}$ years and the exponent of the star formation rate $n = 1$. For more details see \cite{pereira}.
\newpage

\newpage
\bigskip
\bigskip

\textbf{4.2 OGC{\_}lib Usage}
\bigskip
\bigskip

In Python a package is basically a hierarchical file directory structure which defines a environment that consists of modules. This directory contains a file named {\_}{\_}ini{\_}{\_}.py that identifies the directory as a package \cite{pybook}. In the context of this work, this means that if a user does not want to use the GUI of OGCOSMO, he (or she) can use only determined methods by importing its specific class from a module within OGC{\_}lib. For example, an user want to call the method age from the class Cosmo of the module OGC{\_}cosmo:   
\bigskip

$[1]>>>$ from OGC{\_}lib.OGC{\_}cosmo import Cosmo

$[2]>>>$ MyUniverse = Cosmo(0.04,0.24,0.73,0.76,6.0,20.0,'./trabalho')

$[3]>>>$ Age = MyUniverse.age(5)

$[4]>>>$ print` Age = {\%}3.9e' {\%}Age

$[5]$~~~~~~~~~Age = 1.189273236e+09
\bigskip

In line $[1]$ was imported, from OGC{\_}cosmo module within OGC{\_}lib package, the class Cosmo. In line $[2]$  was created a new object (MyUniverse) and  in line $[3]$  was used the method age. The example above show that for  $\Omega_{m} = 0.24$, $\Omega_{b} = 0.04$, $\Omega_{\Lambda} = 0.76$ and $h = 0.73$, the age of the Universe at $z = 5$ is  $1.2\times 10^{9}$ years. The last argument of the class Cosmo  ('./trabalho') is the directory where the data will be saved. 

This result shows that OGC{\_}lib can be used as a framework for fast an easy development of astrophysical and cosmological applications for the study of the cosmic history, within hierarchical scenario for structure formation.

\bigskip
\bigskip

\textbf{5. Final Considerations}

\bigskip
\bigskip

In this work are presented both the  initial stage of the OGCOSMO software and the  OGC{\_}lib package, that have been developed to be an auxiliary tool for reconstruction of the cosmic history. Moreover, it is possible for an user to have access for a specific method of the OGC{\_}lib. This is an important feature to show that this package can be used as a framework for construction of others softwares. An another aspect is that the program can be used as didactic tools in lectures of theory of general relativity and cosmology.

The next steps will be to write  classes to obtain the mass function of SMBHs; research of stochastic  background of gravitational waves (GW) generated by: collapse of stars to black holes, binary systems of SMBH; calculate the signal/noise rate for GW detectors (LISA, LIGO, Decigo, BBO).

Even more, it was showed that the use of PYTHON with FORTRAN for high-level design program is a  powerful tool for development of astrophysical softwares.

\bigskip
\bigskip
{\bf ACKNOWLEDGMENTS: E. S. Pereira would like to thank the Brazilian Agency CAPES for support. O. D. Miranda would like to thank the Brazilian Agency CNPq for partial support (grant 300713/2009-6)

\end{document}